\documentclass[amsmath,pra,showpacs,showkeys]{revtex4}

\usepackage{amsmath}

\usepackage[dvips]{epsfig}
\usepackage{hyperref}
\usepackage{dcolumn}
\usepackage{bm}
\usepackage{mciteplus}

\newcommand{\rem}[1]{}

\begin{document}


\title{Algebraic-matrix calculation of vibrational levels
of triatomic molecules 
}

\author{T. \v{S}edivcov\'a-Uhl\'\i kov\'a$^1$}
\email{Tereza.Sedivcova@gmail.com}
\author{Hewa Y. Abdullah$^{1,2}$}
\author{Nicola Manini$^1$
}
\affiliation{%
$^1$Physics Department and INFM - University of Milan \\
and European Theoretical Spectroscopy Facility\\
Via Celoria 16, 20133 Milano, Italy
\\
$^2$Physics Department, College of Science Education\\
Salahaddidin University, Erbil, Iraq 
}

\date{December 1, 2008}

\begin{abstract}
We introduce an accurate and efficient algebraic technique for 
the computation of the vibrational spectra of triatomic molecules, of both
linear and bent equilibrium geometry.
The full three-dimensional potential energy surface (PES), which can be
based on entirely {\it ab initio} data, is parameterized as a product
Morse-cosine expansion, expressed in bond-angle internal coordinates, and
includes explicit interactions among the local modes.
We describe the stretching degrees of freedom in the framework of a
Morse-type expansion on a suitable algebraic basis, which provides exact
analytical expressions for the elements of a sparse Hamiltonian matrix.
Likewise, we use a cosine power expansion on a spherical harmonics basis
for the bending degree of freedom.
The resulting matrix representation in the product space is very sparse and
vibrational levels and eigenfunctions can be obtained by efficient
diagonalization techniques.
We apply this method to carbonyl sulfide OCS, hydrogen cyanide HCN,
water H$_2$O, and nitrogen dioxide NO$_2$.
When we base our calculations on high-quality PESs tuned to the
experimental data, the computed spectra are in very good agreement with the
observed band origins.
\end{abstract}


\maketitle

\section{Introduction}
The present-day fast development of new spectroscopic instruments and
methods allows us to measure vibrational states with high accuracy even in
the energy region near the molecular dissociation
\cite{ulenikov,urban,voit}.
Characterizing these new experimental data is especially important for
instance, for understanding the dynamics of chemical reactions and for
analyzing spectra from remote regions (e.g., upper atmosphere and/or
interstellar matter), where strong radiation along with low pressure can
stabilize molecules in extraordinarily excited states, sometimes promoting
unusual reactions.
All energy levels depend sensitively on the detailed shape of the molecular
potential energy surface (PES).
Their qualitative and also quantitative description calls for the
development of more and more accurate theoretical approaches
\cite{triatom,dvr3d,morbid,trove,voit,Shirin08}.

Present-day status of the art program codes for the accurate calculation of
vibrational spectra of small molecules have been developed in J. Tennyson's
({\tt TRIATOM \cite{triatom}, DVR3R \cite{dvr3d} ...}) and P. Jensen's
({\tt MORBID \cite{morbid}, TROVE \cite{trove}}) groups.
They are based on different approaches and find best use for different kinds
of small (in particular triatomic) molecules.
{\tt TRIATOM} focuses mainly on van der Waals complexes, and employs
Legendre and Laguerre polynomials as basis functions.
The matrix elements are integrated numerically using a discrete variable
representation (DVR) based on Gauss-Jacobi and Gauss-Laguerre quadrature
for all 3 internal coordinates.
The resulting spectra depend on the DVR grid in a non-variational fashion.
The {\tt MORBID} code is useful mainly for standard rigid molecules.  It
takes an advantage of Morse oscillator basis functions, for which the
matrix elements of pure stretching motion are know analytically.
However, the bending basis functions are evaluated in the framework of a
numerical integration technique, which can be a time-consuming step.
Modern fully variational approaches use a finite basis representation of
the vibrational space: as the {\tt TROVE} implementation in principle
allows to perform variational calculations for general polyatomic molecules
of arbitrary structure.
However in that approach, the kinetic energy operator is represented only
approximately as a power expansion in terms of internal coordinates.
That approximation produces adequately accurate energy levels only rather
low in energy, where the wavefunction has a limited spread around the
molecular equilibrium position, where the power expansion is accurate.

The present work introduces a virtually exact variational method, which
joins the main advantages of (i) simple algebraic forms of matrix elements,
(ii) completeness of the basis set, and (iii) sparseness of the resulting
Hamiltonian matrix.
In concrete, we describe the stretching modes with a formalism based on the
Morse oscillator \cite{bordoni1}, with some similarity to the algebraic
approaches of Ref.~\cite{levine}.
The main advantage of an potential expansion in terms of Morse coordinates
is that one can use a quantum-mechanical basis on which the matrix
representation of the Hamiltonian operator is sparse and can be computed
analytically using algebraic techniques based on generalized step
operators.
For the bending mode we choose a cosine-power expansion of the potential
and a quantum mechanical basis of spherical-harmonic functions.
We then formulate and solve numerically the multi-dimensional problem of
the vibrations of a polyatomic molecule in a product space of the different
degrees of freedom.
The sparse quality of the individual matrices representing the one-dimensional
vibrations and the interaction terms carries forward to the global
matrix representation of the total Hamiltonian on the product Hilbert space.
The eigenvalues and eigenvectors of the resulting sparse matrix can then
be obtained numerically quite efficiently by modern exact diagonalization
tools.
In principle, this method can be applied to molecules composed by any number
of atoms, even though in practice the size of the product Hilbert space
grows exponentially in the number of dimensions, which makes this method,
like all virtually exact methods rather unpractical for molecules with 6 or
more atoms.
In this study we focus on triatomics, but we construct the theory in terms
of basic building blocks usable for successive extensions to the
multi-dimensional potential surfaces of polyatomic molecules, which will be
the subject of future investigation.

The present approach bears some resemblance to Lie algebraic methods based
on a Heisenberg formulation of quantum chemistry (the second quantization
of the Schr\"{o}dinger equation).
Several previous studies \cite{Iachello92,iachello_2002,levine,iachello}
use an algebraic approach where the full Hamiltonian operator is expanded
in powers of $\cal C$, a Casimir operator of a suitable Lie
algebra, for example as follows: $\hat{H}=E_0+\sum_{i=2}^k A_i
[{\cal C}(O(2))]^i$.
The trouble with that kind of expansion is that it maps to an explicit
first-quantized Hamiltonian form involving an intricate mixed potential and
kinetic contributions, including high powers of the momentum operator and
unphysical products of the momentum and position operators.
To avoid such unwanted properties we express separately the potential $V$
and kinetic $T$ energy operators.
This can be done within a convenient scheme of ladder operators defined by
means of a suitable factorization method \cite{bordoni1,bordoni2}.

Section \ref{theory:sec} sketches the model used, including the
Hamiltonian, basis functions for stretching and bending modes, and formulas
for the matrix elements.
Section \ref{results:sec} reports and discusses the vibrational spectra of
OCS, HCN, NO$_2$, and H$_2$O as obtained with the present method.
Section \ref{future:sec} discusses the approximations involved in the
present method and its future extensions.

\section{Theory}
\label{theory:sec}

\begin{figure}
\centerline{\epsfig{file=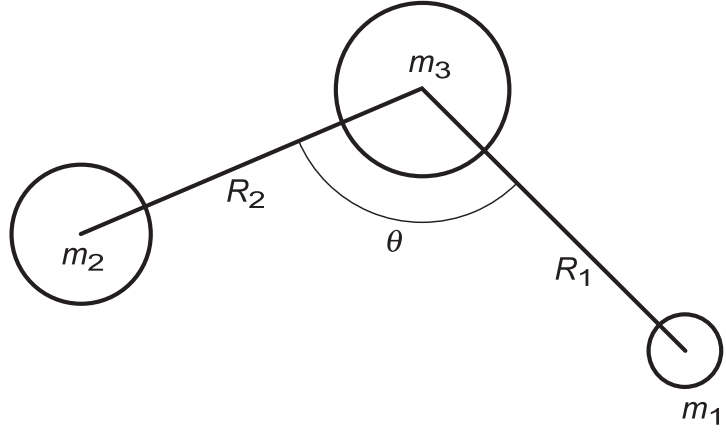,width=95mm,clip=}}
\caption{\label{m3:fig1}
The scheme of internal coordinates used in the present work.
}
\end{figure}

We perform all calculations in the bond-length-angle internal coordinates,
as illustrated in Fig.~\ref{m3:fig1}. This is the most suitable choice of
variables for the potential expansion involving Morse functions.
$R_1$ and $R_2$ represent the bond lengths between the central atom and the
two end atoms, $\theta$ is the bending angle at the central atom.
Given the masses $m_1$ and $m_2$ of the end atoms and $m_3$ of the central
atom, we define the two diatom reduced masses $\mu_1 = m_1m_3/( m_1+m_3)$,
$\mu_2 = m_2m_3/(m_2+m_3)$.
The vibrational Hamiltonian, composed by a kinetic and a potential part,
can be expressed in atomic units ($\hbar=m_e=q_e=1$) as follows:
\begin{equation}\label{ham}
H(R_1,R_2,\theta) = T(R_1,R_2,\theta) + V(R_1,R_2,\theta)
\,.
\end{equation}
The standard expression for the pure vibrational kinetic energy operator in
internal coordinates \cite{carter,carter_86} is
\begin{eqnarray}\nonumber
T(R_1,R_2,\theta) &=&
-{1 \over 2\mu_1}{\partial^2 \over \partial R_1^2}
-{1 \over 2\mu_2}{\partial^2 \over \partial R_2^2}
\\\nonumber
& &
-\frac 12 \left({1 \over \mu_1 R_1^2}+{1 \over \mu_2 R_2^2}
-{2 \cos \theta \over m_3 R_1 R_2}\right)
\left({\partial^2 \over \partial \theta^2} + \cot \theta {\partial\over
  \partial\theta}\right)
\\\label{kinetic}
& &
-{\sin \theta \over m_3 R_1 R_2} {\partial\over \partial\theta}
+{\cos \theta \over m_3}\left({1 \over R_1}{\partial\over \partial R_2}
+{1 \over R_2}{\partial\over \partial R_1} \right)-{\cos \theta \over
  m_3}{\partial^2 \over \partial R_1\partial R_2} 
\\\nonumber
& &
+{\sin \theta \over m_3}{\partial\over \partial\theta}
\left({1 \over R_1}{\partial\over \partial R_2}+{1 \over R_2}
{\partial\over \partial R_1} \right)
-{\cos \theta \over m_3 R_1 R_2}
\\\nonumber
& &
+ \frac 18 \left({1\over \mu_1R_1^2}+ {1\over
\mu_2R_2^2}+{2\over m_3R_1R_2}\right) \frac 1{\cos^2(\theta/2)}\, \Pi_{z}^2
\,.
\end{eqnarray}
This expression is singular at four boundary regions: at $R_i\to 0$, and at
the bending extremes $\theta=0$ and $\theta=\pi$.
While the former constitute no serious practical problem, since the
vibrational motion avoids these regions due to the strongly repulsive
nature of the potential energy function $V$ there, the angular singular
points are easily reached by the motion: in particular $\theta=\pi$ is the
angular minimum energy direction in the case of linear molecules, and this
may require explicit care to obtain a numerically stable algorithm.
Specifically, the $\Pi_{z}$ operator in the final row, defined in
Ref.~\cite{carter_86}, accounts for the rotations of the molecule around an
axis attached to it and coincident with the molecular axis when the
molecules reaches its linear configuration $\theta =180^\circ$: to our
purpose $\Pi_{z}$ can be replaced everywhere with its eigenvalue $m$.
The final term $\propto \Pi_{z}^2$ describes the kinetic-energy
contribution of a rotational degree of freedom of bent molecules (and in
this case we omit it), but it is needed to account for the fourth
vibrational degree of freedom describing the $m=\pm 1$ ``$\Pi$'', $m=\pm 2$
``$\Delta$'', etc....  axially-rotating vibrational excitations of those
molecules such as OCS and HCN which are linear in their equilibrium
geometry.
For simplicity, in the main text we will stick mostly to $m=0$, while the
Appendix shall deal with the general algebra describing an arbitrary
integer value of $m$.

We use the following parameterization of the potential energy surface:
\begin{equation}\label{totpotential}
V(R_1,R_2,\theta) =
\sum_{k_1,k_2,k_3=0}^{k_1+k_2+k_3\leq N_c} a_{k_1k_2k_3} \,
v_1(R_1)^{k_1} \, v_2(R_2)^{k_2} \, u(\theta)^{k_3}
\,.
\end{equation}
This power expansion is realized in terms of Morse-related functions
\begin{equation}\label{vofR}
v_i(R_i)=e^{-\alpha_i(R_i-R_{i\,\rm min})}-1
\,,
\end{equation}
for the stretching degrees of freedom $i=1$ or $2$,
and in terms of trigonometric expressions $u(\theta)=\left[\cos \theta
  -\cos \theta_{\rm min}\right]$ for the bending degree of freedom.
Similar parameterizations are employed e.g., by P. Jensen in the {\tt
  MORBID} code \cite{jensen}, also D. Xie and coworkers use this
Morse-cosine expansion for several molecules, with the parameters first
fixed to fit an {\it ab-initio} PES, and later adjusted to reproduce
spectroscopic experimental data \cite{xie_ocs,xie_h2o}.

As customary, to compute a variational solution of the eigenproblem
corresponding to~$H$, we resort to the expansion of the eigenfunctions of
$H$ on a product basis:
\begin{equation}
\Psi_v(R_1,R_2,\theta)=
\sum_{j_1,j_2,j_3}c_v^{j_1\,j_2\,j_3}\, \Phi_{j_1\,j_2\,j_3}(R_1,R_2,\theta)
\,,
\end{equation}
where $v$ is a complete set of vibrational quantum numbers characterizing
an eigenstate of~$H$, $c_v^{j_1\,j_2\,j_3}$ are yet-to-be-determined
expansion coefficients, and
\begin{equation}
\Phi_{j_1\,j_2\,j_3}(R_1,R_2,\theta)=
\phi_{1\,j_1}(R_1)\phi_{2\,j_2}(R_2)Y_{j_3}(\theta)
\,.
\end{equation}

The 1-dimensional basis functions $\phi_{i\,j_i}(R_i)$ for the stretching
modes were discussed in detail in Refs.~\cite{bordoni1,bordoni2}.
In particular, we use here the generalized quasi-number state basis
\cite{bordoni2,molnar}, defined as follows:
\begin{equation}\label{GQNSB}
\phi_{i\,j_i}(R_i)=\sqrt{\frac{\alpha_i \, j_i!}{\Gamma (2\sigma_i +j_i)}}
\,
y_i^{\sigma_i} \, e^{-\frac{y_i}{2}}\, L_{j_i}^{2\sigma_i -1}(y_i)
\,,\qquad
 j_i=0,~1,~2,\dots\,, \qquad i=1 \hbox{ or } 2\,,
\end{equation}
with
\begin{equation}\label{ydef2}
y_i=y_i(R_i)=(2s_i+1)\, e^{-\alpha_i(R_i-R_{i\,\rm min})}\,.
\end{equation}
Here $L_n^{\rho}$ are generalized Laguerre polynomials, $\Gamma$ is the
standard Gamma function generalization of the factorial \cite{arfken} and
$s_i$ and $\sigma_i$ are suitable positive parameters, whose value we
discuss below.

For the angular variable we use the following basis derived from the
spherical harmonics
\begin{equation}\label{angbasis}
Y_{l}(\theta) = \sqrt{2\pi}\, Y_{l\,0}(\theta,\varphi) = N_l
\,P_l(\cos\theta)
\,, \qquad
{\rm with}\ N_l = \left(l+\frac 12\right)^{1/2}
\end{equation}
[$P_l(\cos \theta)$ are standard Legendre polynomials $P_l(z)\equiv \frac
  d{dz^l} (z^2-1)^l/(2^l\,l!)$], which form a convenient orthonormal basis
set over the range $0\leq\theta\leq\pi$, in the measure $d\cos\theta$.
This basis was used successfully by Carter and Handy \cite{carter}.
In that work the authors evaluated the matrix-element integrals numerically
by Gauss-Legendre quadrature, while here we use analytical expressions
detailed below.

As the Hamiltonian is a sum of terms, its matrix elements are also
expressed as a sum of individual terms, and each one of them is expressed
as a product of operators acting on the $R_1$, $R_2$ and $\theta$ variables.
Accordingly, the matrix elements of $H$ are computed as sums of
products of terms, each of which refers to one oscillator individually.
For example, the matrix elements of one of the kinetic contributions in
Eq.~\eqref{kinetic} are evaluated as
\begin{equation}
\langle \Phi_{j_1\,j_2\,j_3}|
-{\cos \theta \over m_3 R_1 R_2}
|\Phi_{j_1'\,j_2'\,j_3'}\rangle
=
-\frac1{m_3}
\langle\phi_{j_1}|\frac 1{R_1}|\phi_{j_1'}\rangle\,
\langle\phi_{j_2}|\frac 1{R_2}|\phi_{j_2'}\rangle\,
\langle Y_{j_3}|\cos \theta |Y_{j_3'}\rangle
\,.
\end{equation}
Each of these 1-dimensional matrix elements is then evaluated using the
algebraic methods described below.

\subsection{Stretching-coordinate matrix elements}

\begin{figure}
\centerline{\epsfig{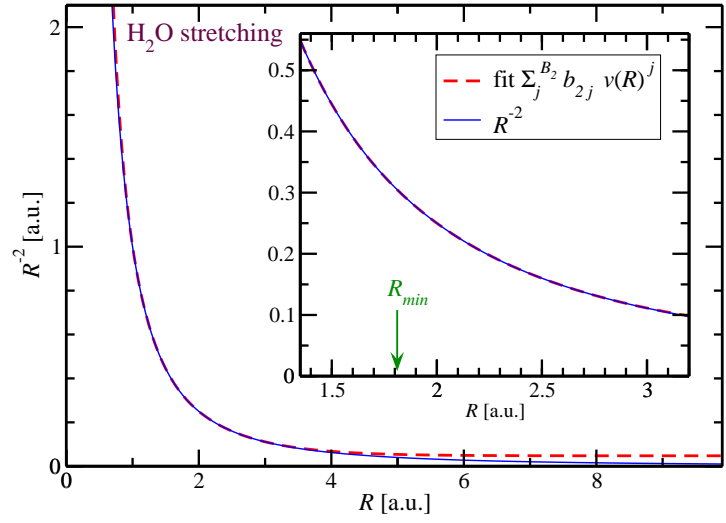}}
\caption{\label{rm2:fig}
Comparison between one of the $R_i^{-2}$ function of Eq.~\eqref{kinetic},
solid line, indicated simply as $R^{-2}$, and its fit in terms of
Morse-related functions, Eq.~\eqref{morsefit}, dashed line.
In this calculation, relevant for the H$_2$O stretching coordinate $R$,
the parameters $\alpha\simeq1.18\, a_0^{-1}$, $R_{\rm min}\simeq 1.81\,a_0$,
and the expansion extends to order $B_2=5$.
The coefficients $b_{2\,j}$ are adjusted to obtain a best fit of equally
spaced points in the $1.35$ to $3.2\, a_0$ range of $R$ highlighted in the
inset.
}
\end{figure}

As a first step, all $R_i$-dependent terms in the Hamiltonian of
Eq.~\eqref{ham} must be expressed as functions of the Morse variable
$y_i(R_i)$ of Eq.~\eqref{ydef2}.
In other words, the $v_i(R_i)^{k_i}$ potential terms in
Eq.~\eqref{totpotential} and the $R_i^{-1}$ and $R_i^{-2}$ terms in the
kinetic energy operator Eq.~\eqref{kinetic} must be expressed in terms of
the corresponding $y_i(R_i)$.
This is accomplished immediately for the potential terms \cite{bordoni2},
by choosing for each mode $i$ the same value of $\alpha_i$ in the potential
expansion terms \eqref{vofR} and in the definition of $y_i(R_i)$,
Eq.~\eqref{ydef2}, so that $v_i\equiv y_i/(2 s_i+1) -1$.
The matrix elements of $R_i^{-1}$ and $R_i^{-2}$ on the basis of
Eq.~\eqref{GQNSB} could be computed exactly by numerical integration, but
that approach would be contrary to the general spirit of the present method
and very inefficient, since that procedure would produce a non-sparse matrix.
We prefer to fit the kinetic terms with a sum of powers of $v_i$:
\begin{equation}\label{morsefit}
 R_i^{-p}\simeq \sum_j^{B_{p\,i}} b_{p\,i\,j} \, v_i(R_i)^j,
\qquad p=1,2
\,.
\end{equation}
An example of the quality of such a fit is illustrated in
Fig.~\ref{rm2:fig}.
The fit targets a reasonably wide region around the equilibrium point
$R_{i\,\rm min}$, where it is extremely accurate, but it deteriorates
especially in the large-$R_i$ dissociation region.
The fitting function could be forced to reach the same large-$R_i$ limit as
$R_i^{-2}$, but then the fit would deviate much more in the most important
region near the minimum.
Accordingly, we prefer to accept this deviation near dissociation, which
causes negligible numerical error to the final spectra.
The fitted $R_i$-ranges and the best-fit coefficients $ b_{1\,i\,j}$ and
$b_{2\,i\,j}$ of $R_i^{-1}$ and $R_i^{-2}$ for OCS, HCN, NO$_2$ and H$_2$O
are reported in the Table A of the supporting information \cite{epaps}.
Root mean square (RMS) deviations ranging from $3$ to $20$~cm$^{-1}$ are
obtained for the fits based on Eq.~\eqref{morsefit}, of the type
exemplified in Fig.~\ref{rm2:fig}.
In practice these deviations affect the computed vibrational band origins
at a completely negligible level.

The full details of the formalism to construct the analytical matrix
elements of the functions $v_i(R_i)$ of Eq.~\eqref{vofR} are reported
elsewhere \cite{bordoni1,lemus_morse}.
Here we just note that the idea underlying the algebraic approach is
ultimately related to supersymmetry, which provides several analytical
relations for a class of exactly-solvable problems, including the
Schr\"{o}dinger equation for the Morse potential.
We only report the basic expression of the matrix elements of the
exponential function and of the first derivative term:
\begin{eqnarray}
\langle \phi_{i\,j} |e^{-\alpha_i(R_i-R_{i \,\rm min})} |
\phi_{i\,j'} \rangle
&=&
{-C_{i\,j'}\,\delta_{j,j'-1}
+2(\sigma_i +j')\,\delta_{j,j'}
-C_{i\,j}\,\delta_{j,j'+1}\over 2s_i +1}\,,
\\
\langle \phi_{i\,j} |\frac{\partial}{\partial R_i} |
\phi_{i\,j'} \rangle
&=&
\frac{\alpha_i}2 \left(C_{i\,j'}\,\delta_{j,j'-1}
-C_{i\,j}\,\delta_{j,j'+1}\right)
,
\end{eqnarray}
where $C_{i\,j}=\sqrt{j(j+2\sigma_i -1)}$.
Based on these expressions, all matrix elements of every stretching term in
the potential expansion Eq.~\eqref{totpotential} and in the kinetic terms
expressed as in Eq.~\eqref{morsefit} can be computed exactly.
The matrix representation of $v_i(R_i)^{k_i}$ is $(2k_i+1)$-band
diagonal.

The computation of the spectrum of one pure-stretching mode through the
exact diagonalization of algebraic matrices can be made substantially more
efficient by choosing a properly adapted quantum-mechanical basis,
specifically tuned to the molecular potential \cite{bordoni2}.
A better convergence of the calculation of a single oscillator improves
the convergence of the full 3-dimensional spectral calculation for the
triatomic molecule.
For example, for the first stretching oscillator we consider
\begin{equation}\label{H1}
H_1(R_1) =
-{1 \over 2\mu_1}{\partial^2 \over \partial R_1^2}
+
\sum_{k_1=0}^{N_c} a_{k_1\,0\,0} \,
v_1(R_1)^{k_1} 
\,.
\end{equation}
The shape of the basis wavefunctions \eqref{GQNSB} is tuned by four parameters
$R_{1\,\rm min}$, $\alpha_1$, $s_1$ and $\sigma_1$.
We fix $R_{1\,\rm min}$ to the relevant equilibrium bond length, and in
order to preserve the matrix sparseness we set $\alpha_1$ to the value used
in the  definition of $v_1(R_1)$.
By tuning the remaining $s_1$ and $\sigma_1$ one has a sufficient basis
flexibility to improve substantially the numerical convergence of the
single-oscillator problem \cite{bordoni2} Eq.~\eqref{H1}.
We optimize the parameters $s_i$ and $\sigma_i$ by the minimization of the
sum of the $N_b$ lowest bound-state eigenenergies of the dimer, as defined
in equation (25) of Ref.~\cite{bordoni2}.
We optimize the energies of $N_b=20$ stretching bound states, which
cover and far exceed the spectral range of interest.
To get sub-cm$^{-1}$ accuracy, a number of basis functions $N_{{\rm
    d}\,i}\simeq 30$ in the one-oscillator basis are usually sufficient.
The values of the final optimized parameters $s_i$ and $\sigma_i$ adopted
in all calculations are collected in Table~\ref{basispar:tab}.
In fact, when $N_{{\rm d}\,i}$ is large enough the spectral accuracy
depends only weakly on the value of the basis parameters $s_i$ and
$\sigma_i$, so that the reported values are not particularly critical.

\subsection{Bending-angle matrix elements}

On the spherical harmonics basis Eq.~\eqref{angbasis}, it is straightforward to
express the matrix elements of powers of the $\cos\theta$ function.
The bending-angle dependence of the potential energy surface is then
conveniently fitted to powers of $(\cos\theta - \cos\theta_{\rm min})$.
In this basis, the bending potential matrix is
sparse and can be expressed analytically.
To evaluate the matrix elements of powers of $z=\cos\theta$, we use the
following relations \cite{arfken}:
\begin{eqnarray}\label{zmatel}
\langle Y_{l}|z|Y_{j} \rangle&=&
 {l\over \sqrt{(2j+1)(2l+1)}} \,\delta_{l,j+1}
+{j\over \sqrt{(2j+1)(2l+1)}} \,\delta_{l,j-1}
\,,
\\
\langle Y_{l}|z^2|Y_{j}\rangle&=&
 {2j^2+2j-1 \over (2j-1)(2j+3)}
\,\delta_{l,j}
 \\\nonumber
&&+{l(j+1)\over (j+l+1)\sqrt{(2j+1)(2l+1)}}
\,\delta_{l,j+2}
+{j(l+1)\over (j+l+1)\sqrt{(2j+1)(2l+1)}}
\,\delta_{l,j-2}
\,,
\end{eqnarray}
and in general, recursively
\begin{equation}\label{recurszed}
\langle Y_{l}|z^{k_3}|Y_{j}\rangle=
{j+1 \over \sqrt{(2j+1)(2j+3)}} \, \langle Y_{l}|z^{k_3-1}|Y_{j+1}\rangle
+{j \over \sqrt{(2j+1)(2j-1)}}\, \langle Y_{l}|z^{k_3-1}|Y_{j-1}\rangle
\,.
\end{equation}
Like for stretching matrix elements, the matrix representing $
u(\theta)^{k_3}$ is $(2{k_3}+1)$-band diagonal.

The derivation of the kinetic bending matrix elements is described in
Appendix \ref{app:ang}.
The final result is the following simple tridiagonal expression:
\begin{eqnarray}\label{angfinmatel}
\langle Y_{l}|T_{\rm bend}|Y_{j} \rangle&=&
\frac 12 \left({1 \over \mu_1 R_1^2}+{1 \over \mu_2 R_2^2}\right)j(j+1)
\delta_{l,j}
\\ \nonumber
&&-\frac 1 {m_3 R_1 R_2} \,\frac{1}{\sqrt{(2l+1)(2j+1)}} 
\left(
l^3 \, \delta_{l,j+1} +
j^3 \, \delta_{l,j-1}
\right)
.
\end{eqnarray}

The basis functions Eq.~\eqref{angbasis} are independent of the shape of
the bending potential, which means that, for bent molecules in particular,
the number of basis functions required by a given degree of convergence
could become unreasonably large.
Unfortunately, there is no way of optimizing the individual functions to
any given angular potential, as we do for the stretching functions.
One way to overcome this problem is to replace the angular functions
\eqref{angbasis} with suitably optimized linear combinations thereof, for
example, obtained as low-energy eigenstates of a purely bending
1-dimensional problem based on $T_{\rm bend}$ plus the $a_{0\,0\,k_3}$ part
of the potential expansion \eqref{totpotential}, as was suggested e.g., in
Ref.~\cite{carter} and Ref.s therein.
Such kind of approach can lead to significant basis size reduction, but
also, unfortunately, to entirely nonzero angular matrix elements,
eventually leading to a less sparse total matrix of $H$.

\section{Results}
\label{results:sec}

We come to illustrate the application of the proposed algebraic method to
the calculation of the vibrational spectra of real triatomic molecules.
We target mainly the purely vibrational energy levels ($J=0$).
As trial systems we select two linear molecules, OCS and HCN, and two bent
ones, NO$_2$ and H$_2$O.
They are all well studied by experimental as well as theoretical techniques
which provide accurate spectra to compare with.
In particular, for OCS, NO$_2$ and H$_2$O the literature offers realistic
PESs parameterized in the form \eqref{totpotential} suitable for our method
\cite{xie_ocs,xie_no2,xie_h2o}.
Accordingly, we have taken the PES function parameters $a_{k_1k_2k_3}$ for
OCS, NO$_2$ and H$_2$O exactly as in the referred publications
\cite{sign:note}.
For HCN, we use the potential surface provided by Ref.~\cite{mourik}
to fit an expansion of the type \eqref{totpotential}.
All these PESs are determinated starting from {\it ab initio} points, but
then the function parameters $a_{k_1k_2k_3}$ are adjusted so that the
deviation between the calculated vibrational levels and the observed
spectra is minimized.
In the energy region considered, these PESs are accurate enough to allow
one to predict even highly excited vibrational levels with great precision.

The values of the Morse exponential parameters $\alpha_i$ and equilibrium
positions $R_{i\,\rm min}$ are collected in Table~\ref{potparam:tab}.
We fit the algebraic potential parameters $a_{k_1k_2k_3}$ for HCN to 737
points obtained using the PES function as given in Ref.~\cite{mourik},
restricted to the HCN side of the HCN-HNC isomerization transition.
The angular range is $\theta=180^\circ$ to $90^{\circ}$, the coordinate
ranges are $R_{\rm HC}/a_0=1.4$ to $4.4$ and $R_{\rm CN}/a_0=1.6$ to $3.0$,
with energy up to 45\,000\,cm$^{-1}$.
Table B in the supporting information \cite{epaps} reports the resulting
best fit parameters $a_{k_1k_2k_3}$, which produce a RMS deviation of
29~cm$^{-1}$.

As the bending frequency is usually smaller than stretching, and as the
bending basis \eqref{angbasis} cannot be optimized to the problem at hand,
it is generally necessary to include a larger number $N_{{\rm d}\,3}$ of
bending basis states.
The minimum size $N_{\rm d}=N_{{\rm d}\,1}\,N_{{\rm d}\,2}\,N_{{\rm d}\,3}$
of the matrix for the total Hamiltonian \eqref{ham} which needs to be
diagonalized for a very accurate and reasonably stable spectrum is
eventually very moderate, of order $N_{\rm d}\sim 10^4$.
The nonzero matrix elements of $H$ are of the order of $\sim \left[\frac 23
  \max (N_c,B_{1\,1},B_{1\,2},B_{2\,1},B_{2\,2})+1\right]^3$ in each row of
the global matrix.
For $B_{k\,i}=5\geq N_c$, as in most calculation of the present work, this
yields approximately 80 nonzero matrix elements per row, i.e.\ a very
sparse matrix to store and diagonalize.

Once the full matrix is constructed and stored, it is diagonalized using
the Jacobi-Davidson method provided by the {\tt PRIMME}
package \cite{prime1,prime2} for sparse matrices.
Table~\ref{RMS:tab} summarizes the quality of the computed spectra, in
terms of root mean squared (RMS) deviations between the observed and
calculated vibrational energy levels.
These levels cover an energy range extending from the ground state up to
several thousand wavenumbers, as indicated in Table~\ref{RMS:tab}.
The good agreement with experiment shown by the small RMS deviations
indicates both the high quality of the considered PES parameterizations and
the satisfactory accuracy of the employed method.
The complete spectral levels obtained by means of the present approach and
their detailed comparison to experimental data for OCS, HCN, NO$_2$ and
H$_2$O are collected in Tables C, D, E, F in the supporting information
\cite{epaps}.
For OCS we compute and compare to experiment also numerous $\Pi$ ($m=\pm 1
$) states.

For the molecules for which spectra were computed before based on the same
PES (but with a different approach to the solution of the
quantum-mechanical problem) we obtain an essentially equivalent accuracy of
the spectra, as shown by the similar RMS deviations reported in
Table~\ref{RMS:tab}.
Basically all discrepancies with experiment are to be attributed to the
lack of accuracy of the adopted PES.
Table D in the supporting information \cite{epaps} collects also a few
predicted energy levels of HCN in the $7000$ to $12000$~cm$^{-1}$ spectral
range.
These calculated values are also in good agreement with previous
calculations by Mourik {\it et al.}
Table F in the supporting information \cite{epaps} reports also a few
predicted levels for H$_2$O.
%

The level assignments in terms of local vibrations is not always
straightforward.
The wave function of each vibrational state can be analyzed and it always
results in a complicated admixture of excitations of all three local
vibrational modes.
For instance, the OCS vibrational excitation at $E_v=2937.2$~cm$^{-1}$
consists mainly (35\%) of $v=(1\,4\, 0)$ plus 21\% $v=(0\, 6\, 0)$, plus
other minor components, which confirms the traditional assignment $(1\,4\,
0)$ reported in Table C.
However, for example the largest local component, $v=(0\,8\,0)$, of the OCS
vibrational excitation near 3990~cm$^{-1}$ is 26\%, while the component on
its standard local assignment $v=(1\,6\,0)$ is smaller (18\%).

\section{Discussion and conclusions}
\label{future:sec}

In the present work, we demonstrate the possibility to compute the
vibrational spectrum of an arbitrary triatomic molecule based on its PES
and using algebraic techniques for the analytical determination of the
matrix elements.
The advantages of this method include simple formulas for the matrix
elements, moderate total Hilbert space size, sparseness of the Hamiltonian
matrix, all cooperating to a fast and efficient diagonalization.
%
%
The generalization of this method to four-atomic and larger molecules is in
principle straightforward, and in these higher-dimensional contexts the
reduced basis size for each degree of freedom is even more crucial.
These advantages make this method a very promising tool for the analysis of
the vibrational levels of small molecules.

The input of the present approach includes only the atomic masses and a
reasonably dense numerical sampling of the PES: this is in principle within
reach of {\it ab-initio} electronic structure quantum chemical
calculations.
We did carry out a calculation of the spectrum of HCN based on the {\it
  ab-initio} PES generated by a large set of DFT-LDA (density functional
theory in the local-density approximation) calculations.
Due to the known drawbacks of the LDA, the resulting spectrum shows
deviations from the experimental levels of a few hundred cm$^{-1}$, but it
proves the possibility to compute even highly excited vibrational molecular
states entirely from first principles, with no parameters adjusted to the
observed spectroscopical data.
When highly accurate quantum chemical methods are applied to a relatively
fine and extensive determination of the PES of a triatomic molecule, it
will be possible to obtain much better predictive power for a calculation
entirely free from any experimental input.

If the PES is know to a high degree of accuracy even in the energy region
near and slightly above dissociation, the present method can take advantage
of the possibility to extend the basis set beyond the number of bound
states to describe reliably weakly-bound pre-dissociation states in the
quasi continuum.
In the future, the present algebraic approach could be extended to the
study of resonances in the continuum, bound-to-free transitions in infrared
absorption, Franck-Condon processes and in principle even atom-molecule and
molecule-molecule collisions.
Other perspective applications include the area of non-rigid molecules (van
der Waals complexes, quasi-linear molecules) \cite{Iachello92} and
potentials with many minima such as those occurring in torsional
oscillations \cite{iachello_2002}.
Note finally that a reliable description of the quantum vibrational
dynamics could allow us to exploit the manageable Hamiltonian structure
produced by the present method to the study of intramolecular
vibrational-energy redistribution.

\acknowledgments

The authors wish to thank A. Bordoni for useful discussion, T. Mourik and
J. Tennyson for kindly providing the computer procedure for generating the
pointwise HCN potential from their analytical function \cite{mourik}, and
the supercomputing centers Cineca (Italy) and Metacentrum (Czech Republic)
for computer time allocation.
This work was supported by the European Union through e-I3 ETSF project
(INFRA-211956) and NoE Nanoquanta (NMP4-CT-2004-500198).
H.Y.A. acknowledges financial support by the Italian Ministry of Foreign
Affairs support and cooperation programme with the Iraqi scientific and
university system, managed by the Landau Network and Centro Volta, Como,
Italy.

\vskip 1cm

{\bf Supporting Information Available:}
Document No. ... contains the best-fit parameters describing the PES of HCN
according to Eq.~\eqref{totpotential}.
This document also reports the $b_{p\,i\,k}$ best-fit parameters of
Eq.~\eqref{morsefit} for the $R_i^{-p}$ functions.
This document also lists detailed tables of the computed vibrational band
origins compared to experimental data.
For HCN and H$_2$O, a comparison is also made with spectral data calculated
using different approaches.
This material is available free of charge via the Internet at
http://pubs.acs.org .

\appendix
\section{Angular matrix elements}\label{app:ang}

The kinetic energy operator, Eq.~\eqref{kinetic}, contains terms
proportional to $\cot \theta$ and to $[\cos(\theta/2)]^{-2}$, which are
singular at the boundary points $\theta=0$ and $\pi$.
In fact $\theta=\pi$ is an especially important configuration for linear
molecules, and it should then be described smoothly.
Luckily, since the integration implied in the matrix elements is performed
over the variable $z=\cos\theta$, this divergence is not a problem, and all
relevant matrix elements can be computed analytically using several
properties of the spherical harmonics.
For brevity, we replace the $\theta$-independent kinetic coefficients with
the shorthand
\begin{eqnarray}
a_1&=&-\frac 12 \left({1 \over \mu_1 R_1^2}+{1 \over \mu_2 R_2^2}\right) \\
a_2&=&{1 \over m_3 R_1 R_2}
\,,
\end{eqnarray}
and define the pure bending part of $T$ as
\begin{eqnarray}
T_{\rm bend}(\theta) &=&
\left(a_1 +a_2 \cos \theta \right)
\left({\partial^2 \over \partial \theta^2} + \cot \theta {\partial\over \partial\theta}\right)
-a_2 \left( \sin \theta {\partial\over \partial\theta} +\cos \theta \right)
\\\nonumber
&&+ \frac {a_2 -a_1}2 \,  \frac 1{1+\cos\theta}\, \Pi_{z}^2
\,.
\end{eqnarray}

In terms of $z=\cos \theta$, 
${\partial\over \partial\theta} \rightarrow
 -\sin \theta {\partial\over \partial \cos \theta}
=-(1-z^2)^{1/2}\, {\partial\over \partial z}
$, and 
${\partial^2 \over \partial \theta^2} \rightarrow
\sin^2 \theta  {\partial^2 \over \partial \cos^2 \theta}
 - \cos \theta {\partial\over \partial \cos \theta}
=
(1-z^2) \, {\partial^2 \over \partial z^2 } - z {\partial\over \partial z}
$,
the angular kinetic energy is conveniently decomposed as
\begin{eqnarray}
T_{\rm bend}(z)=&
a_1(1-z^2){\partial^2 \over \partial z^2}&\rightarrow \ T_1\nonumber\\
&+\,a_2z(1-z^2){\partial^2 \over \partial z^2}&\rightarrow \ T_2\nonumber\\
&+\,a_1\left(-2z {\partial \over \partial z}\right) &\rightarrow \ T_3\nonumber\\
&+\,a_2\left(-2z^2 {\partial \over \partial z}\right) &\rightarrow \ T_4\nonumber\\
&+\,a_2(1-z^2){\partial \over \partial z} &\rightarrow \ T_5\nonumber\\
&-\,a_2 \,z  &\rightarrow \ T_6 \nonumber\\
&+\,\frac {a_2 -a_1}2 \,  \frac 1{1+z}\, m^2
& \rightarrow \ T_7
\,.
\end{eqnarray}

We derive the matrix elements of $T_{\rm bend}$ in the general case of
arbitrary integer $m$.
Because for the $m\neq 0$ ``$\Pi$'', ``$\Delta$'', etc.\ vibrational
excitations of linear molecules, the $T_7$ term is singular at
$\theta=180^\circ$, all its matrix elements in the $Y_{l0}$ basis diverge:
for general $m$ we need then to consider the natural extension of the basis
of Eq.~\eqref{angbasis}, namely the one provided by the $Y_{l\,m}$
spherical harmonics:
\begin{equation}\label{angbasisnonsigma}
Y_l^m(\theta) = \sqrt{2\pi}\,Y_{l\,m}(\theta,0) =
N_{l\,m} \,P_l^m(\cos\theta), \qquad
{\rm with}\ N_{l\,m} = \left[\frac{(2l+1)(l-|m|)!}{2(l+|m|)!}\right]^{1/2}
,
\end{equation}
for $l\geq |m|$.
For compactness, in the following we write the matrix elements in terms of
associated Legendre polynomials $P_l^m(z)\equiv (-1)^{(|m|+m)/2}\,
(1-z^2)^{|m|/2} \, dP_l(z)/dz^{|m|}$, i.e.\ we omit the normalization
factor $N_{l\,m}$, and we use the notation ${P_l^m}'=\frac d{dz} P_l^m$.
In the derivation of the $T_{\rm bend}$ matrix elements, we make use of the
following identities for the associated Legendre polynomials:
\begin{eqnarray}\label{generalrecurrency}
(j+1-|m|)P_{j+1}^m(z)&=&(2j+1)zP_j(z)-(j+|m|) P_{j-1}^m(z),\\
(1-z^2){P_j^m}'(z)&=&-jzP_j^m(z)+(j+|m|) P_{j-1}^m(z) \label{p1}, \\
(1-z^2){P_j^m}''(z)&=&
2z{P_j^m}'(z)-\left[j(j+1)-\frac{m^2}{1-z^2}\right]P_j^m (z) \label{p2}
\,.
\end{eqnarray}
As a direct consequence of Eq.~\eqref{generalrecurrency},
Eq.~\eqref{zmatel} generalizes to
\begin{equation}\label{zmatelmneq0}
\langle Y_l^m|z|Y_j^m \rangle
=
\frac{N_{j\,m}}{N_{l\,m}} \left(
 \frac{l-|m|}{2j+1} \,\delta_{l,j+1}
+\frac{j+|m|}{2j+1} \,\delta_{l,j-1}
\right)
,
\end{equation}
and the recursive relation \eqref{recurszed} generalizes to
\begin{equation}
\langle Y_l^m|z^{k_3}|Y_j^m\rangle=
\frac{N_{j\,m}}{2 j+1} \left(
\frac{j+1-|m|}{N_{j+1\,m}} \, \langle Y_l^m|z^{k_3-1}|Y_{j+1}^m\rangle
+\frac{j+|m|}{N_{j-1\,m}}  \, \langle Y_{l}|z^{k_3-1}|Y_{j-1}\rangle
\right)
.
\end{equation}
These relations are also useful for the calculation of the potential matrix
elements.

The individual kinetic terms are then expressed as
\begin{eqnarray}
\langle P_l^m|T_1|P_j^m \rangle&=&
a_1\langle P_l^m|(1-z^2)|{P_j^m}''\rangle 
\\\nonumber
&=& a_1\langle P_l^m|2z|{P_j^m}'\rangle 
- a_1 j(j+1)\langle P_l^m|P_j^m \rangle 
+ a_1 m^2 \langle P_l^m|\frac 1{1-z^2}|P_j^m \rangle 
\\
\langle P_l^m|T_2|P_j^m \rangle&=&
a_2\langle  P_l^m|z(1-z^2)|{P_j^m}''\rangle 
\\\nonumber
&=& a_2\langle P_l^m|2z^2|{P_j^m}'\rangle 
-a_2j(j+1)\langle P_l^m|z|P_j^m \rangle 
+ a_2 m^2 \langle P_l^m|\frac z{1-z^2}|P_j^m \rangle 
\\
\langle P_l^m|T_3|P_j^m \rangle&=&-a_1\langle P_l^m|2z|{P_j^m}' \rangle 
\\
\langle P_l^m|T_4|P_j^m \rangle&=&-a_2\langle P_l^m|2z^2|{P_j^m}' \rangle 
\\
\langle P_l^m|T_5|P_j^m \rangle&=&
a_2\langle P_l^m |(1-z^2)|{P_j^m}' \rangle 
= -a_2j\langle P_l^m|z|P_j^m \rangle
 + a_2(j+|m|)\langle P_l^m|P_{j-1}^m \rangle 
\\
\langle P_l^m|T_6|P_j^m \rangle&=&-a_2\langle P_l^m|z|P_j^m \rangle
\\
\langle P_l^m|T_7|P_j^m \rangle&=&
\frac {a_2 -a_1}2\, m^2 \, \langle P_l^m| \frac 1{1+z}|P_j^m \rangle
\,.
\end{eqnarray}
By collecting everything together and simplifying we obtain
\begin{eqnarray}
\langle P_l^m|T_{\rm bend}|P_j^m \rangle
&=&
-a_1j(j+1)\langle P_l^m|P_j^m \rangle 
-a_2(j+1)^2\langle P_l^m|z|P_j^m \rangle
\\\nonumber
&&+a_2 (j+|m|)\langle P_l^m|P_{j-1}^m \rangle
+\frac {a_1 +a_2}2\, m^2 \, \langle P_l^m| \frac 1{1-z}|P_j^m \rangle
\,.
\end{eqnarray}
Accordingly, the bending kinetic matrix elements in the normalized basis
$Y_l^m$ are
\begin{eqnarray}\nonumber
\langle Y_l^m|T_{\rm bend}|Y_j^m \rangle&=&
-\,a_1\, j(j+1)\,\delta_{lj}
-a_2\, (j+1)^2 \, \langle Y_l|z|Y_j \rangle 
\\\nonumber
&&+\,a_2 \, (j+|m|) \frac{N_{j\,m}}{N_{j-1\,m}}\,\delta_{l,j-1}
+\frac {a_1 +a_2}2\, m^2 \, \langle Y_l^m| \frac 1{1-z}|Y_j^m \rangle
\\\label{angkinmatel}
&=&
-\, a_1 \, j(j+1)\,\delta_{l,j}
 \\\nonumber 
&&-\,\frac {a_2}{\sqrt{(2j+1)(2l+1)}}
\left[
  l^2 \sqrt{l^2-m^2} \,\delta_{l,j+1} 
+ j^2 \sqrt{j^2-m^2} \,\delta_{l,j-1}
 \right]
\\ \nonumber
&&+ \, \frac {a_1 +a_2}2 \, m^2\, \langle Y_l^m|\frac 1{1-z}|Y_j^m \rangle
\,,
\end{eqnarray}
where we have inserted the matrix elements of $z$ from
Eq.~\eqref{zmatelmneq0}, and combined the two non-Hermitian terms
proportional to $a_2$ into a final explicitly Hermitian expression.
The important special case $m=0$ is reported as Eq.~\eqref{angfinmatel} in
the main text.
The matrix element in the final term is
\begin{equation}\label{angnonsparse}
\langle Y_l^m|\frac 1{1-z}|Y_j^m \rangle
=
N_{l\,m} \, N_{j\,m} \,
l_{\rm max}\,[l_{\rm max}+1] \, K_m(l_{\rm max})
\,,
 \qquad
{\rm with}\ l_{\rm max} = \max(l,j)
\,,
\end{equation}
and with
\begin{eqnarray}\nonumber
K_{\pm 1}(l) &=& 1
\\\nonumber
K_{\pm 2}(l) &=& \frac 12 (l^2+l-2)
\\\nonumber
K_{\pm 3}(l) &=& \frac 13 (l^4+2 l^3-7 l^2-8 l+12 )
\,.
\end{eqnarray}
Unfortunately the last term in Eq.~\eqref{angkinmatel}, the one
proportional to $m^2$ and originated by $T_7$ plus parts of $T_1$ and $T_2$,
produces a matrix which, according to Eq.~\eqref{angnonsparse}, is not
sparse on this basis.
This leads to significant numerical overhead in the calculation of the
$m\neq 0$ states, with respect to the $m=0$ ones.


\newpage
\quad

\begin{table}
\caption{\label{basispar:tab}
Optimized parameters $s_i$ and $\sigma_i$, number of target bound states
$N_b$, and size $N_{{\rm d}\,i}$ of the quantum basis for each stretching
oscillator of all molecules studied in the present work.
}
\begin{ruledtabular}
\begin{tabular}{lddcc}
stretching&\multicolumn{2}{c}{optimized}&number of &number of\\
dimer&\multicolumn{2}{c}{parameters}&bound states&basis functions\\
&s&\sigma&$N_b$&$N_{{\rm d}\,i}$\\
\hline
OC&68.&40&20&30 \\
CS&76.&52&20&30 \\
\hline
HC&30.&0.20&20&40 \\
CN&74.&46&20&30 \\
\hline
OH&26.&0.01 &20&30 \\
\hline
NO&33.&5.04&20&30 \\
\end{tabular}
\end{ruledtabular}
\end{table}

\begin{table}
\caption{\label{potparam:tab}
The values of several relevant potential parameters for the molecules studied
in the present work.
The HCN parameters are obtained through a best fit to 737 points on the
PES as computed in Ref.~\cite{mourik}.
}
\begin{ruledtabular}
\begin{tabular}{llllllc}
system&$R_{1\,\rm min}\ [a_0]$&$R_{2\,\rm min}\ [a_0]$&$\theta_{\rm
  min}\ [{\rm degree}]$&$\alpha_1\ [a_0^{-1}]$&$\alpha_2\ [a_0^{-1}]$&Reference\\
\hline
OCS&2.1849 (OC)&2.9506 (CS)&180.0&1.2382&1.0318&\cite{xie_ocs}\\
HCN&2.0135 (HC)&2.1793 (CN)&180.0&0.9727&1.2290&\cite{mourik}\\
NO$_2$&2.2435&2.2435&133.767&1.6853&1.6853&\cite{xie_no2}\\
H$_2$O&1.8112&1.8112&104.440&1.1769&1.1769 &\cite{xie_h2o}\\
\end{tabular}
\end{ruledtabular}
\end{table}

\begin{table}
\caption{\label{RMS:tab}
Basis size for the individual oscillators $N_{{\rm d}\,i}$, total basis
size $N_{\rm d}$, and standard deviations of the computed spectra to the
available experimental vibrational levels and to previous calculations of
the same levels.
}
\begin{ruledtabular}
\begin{tabular}{lccccr@{\quad}r@{\quad}cr}
system&\multicolumn{4}{c}{number of}&num.\ of&up to&\multicolumn{2}{c}{RMS deviation}\\
&\multicolumn{4}{c}{basis function}&energy&energy&obs.-cal&obs.-Ref.\\
&$N_{\rm d\,1}$&$N_{\rm d\,2}$&$N_{\rm d\,3}$&$N_{\rm d}$&levels&[cm$^{-1}$]&[cm$^{-1}$]&[cm$^{-1}$]\\
\hline
OCS&30 &30 &50 &45$\times 10^{3}$ &145&8057& 0.31& 0.26 \cite{xie_ocs}\\
HCN&40&30&50&60$\times 10^{3}$&34&12389&14.&12. \cite{mourik}\\
NO$_2$&30&30&55&49.5$\times 10^{3}$&143&8979&1.7&2.1 \cite{xie_no2}\\
H$_2$O&30&30&50&45$\times 10^{3}$&69&21247&5.6&1.2 \cite{xie_h2o}\\
\end{tabular}
\end{ruledtabular}
\end{table}

\end{document}